\documentclass[bibnotes,showpacs,twocolumn,aps,pra,a4paper]{revtex4}

\usepackage{color,epsfig,rotating,graphicx,subfigure}
\usepackage{amsmath,amssymb,amscd}

\usepackage[latin1]{inputenc}
\usepackage[english]{babel}

\usepackage{dsfont}
\usepackage{textcomp}

\renewcommand{\Re}{{\rm Re}}

\newcommand{\ri}{{\rm i}}
\newcommand{\re}{{\rm e}}
\newcommand{\rd}{{\rm d}}

\newcommand{\rs}{{\rm s}}
\newcommand{\rp}{{\rm p}}

\newcommand{\kb}{k_{\rm B}}

\newcommand{\kB}{k_{\rm B}}

\begin{document}

%
%
\title{Fluctuations of radiative heat exchange between two bodies}
\author{S.-A. Biehs$^{1}$}
\email{s.age.biehs@uni-oldenburg.de}
\affiliation{$^1$ Institut f\"{u}r Physik, Carl von Ossietzky Universit\"{a}t, D-26111 Oldenburg, Germany}
\author{P. Ben-Abdallah$^{2,3}$}
\email{pba@institutoptique.fr}
\affiliation{$^2$ Laboratoire Charles Fabry,UMR 8501, Institut d'Optique, CNRS, Universit\'{e} Paris-Sud 11,
2, Avenue Augustin Fresnel, 91127 Palaiseau Cedex, France}
\affiliation{$^3$ Universit\'{e} de Sherbrooke, Department of Mechanical Engineering, Sherbrooke, PQ J1K 2R1, Canada.}
%

%
%
\begin{abstract}
We present a theory to describe the fluctuations of nonequilibrium radiative heat transfer between two bodies both in far and near-field regime. As predicted by the blackbody theory, in far field, we show that the variance of radiative heat flux is of the same order of magnitude as its mean value. However, in near-field regime we demonstrate that the  presence of surface polaritons makes this variance more than one order of magnitude larger than the mean flux. We further show that the correlation time of heat flux in this regime is comparable to the relaxation time of heat carriers in each medium. This theory could open the way to an experimental investigation of heat exchanges far from the thermal equilibrium condition.
\end{abstract}

\maketitle

%
%

Nonequilibrium fluctuations in electronic transport~\cite{Akkermans} inside mesoscopic systems have been investigated in detail 
since the beginning of 90's~\cite{Blanter,Beenakker}. In these systems, fluctuations of electric currents were found
to be of the same order of magnitude as their mean value. These fluctuations originate from coherence effects for electronic wavefunctions. 
An analog thermal  behavior is well known for heat flux exchanged in far-field regime between two objects held at two different temperatures. This behavior is a direct consequence of blackbody fluctuations as predicted by Einstein~\cite{Planck, Einstein}. Surprisingly these fluctuations have not been investigated so far at close separation distances. However, in the last two decades it has been shown that the properties of thermal radiation in the near-field regime can radically differ from  that observed in the far field. Indeed in this case the thermal radiation can be quasi-monochromatic~\cite{Eckardt}, polarized~\cite{SetalaEtAl2002} and spatially coherent~\cite{Hesketh}. As the radiative heat flux between two thermalized objects is concerned,  it has been shown within the framework of Rytov's fluctuational electrodynamics~\cite{Rytov} that it can surpass the blackbody limit by orders of magnitude~\cite{PoldervH,LoomisPRB94,Pendry,JoulainSurfSciRep05,VolokitinRMP07,PBA2,BiehsEtAl2010} and strong deviations from the behavior observed in far-field regime have been predicted in a variety of configurations~\cite{RodriguezPRL11,McCauleyPRB12,RodriguezPRB12,MullerarXiv,BimontePRA09,KrugerPRL11,MessinaEurophysLett11,KrugerPRB12,MessinaPRA14,BimontearXiv16,BenAbdallahPRL11,ZhuPRL16,ZhengNanoscale11,MessinaPRL12,BenAbdallahPRL14,MessinaPRB13,BenAbdallahPRL13,NikbakhtJAP14,BenAbdallahAPL06,BenAbdallahPRB08,NikbakhtEPL15,BenAbdallahPRL16,LatellaarXiv16}. Many of these theoretical predictions have been confirmed experimentally  down to few nanometer distances~\cite{Hargreaves,Kittel1,Chen1,Chen2,Chen3,Rousseau,Ottens,Kralik2,Chevrier1,Chevrier2,Pramod1,Pramod2,Lipson,Kittel2}. So far, investigation of radiative heat exchanges between two bodies was limited to the analysis of the statistical average of the Poynting vector {(PV)}~\cite{PoldervH,LoomisPRB94}. To go beyond this first-order theory and to investigate the statistical properties of the near-field thermal radiation it is necessary to determine the high-order moments of fields radiated by the fluctuating sources as well as the heat flux mediated by photon tunneling. The theoretical analyzis of these moments could open, for instance, the way to the investigation of thermodynamical properties of these systems or to the study  of irreversible dynamical processes related to them far from equilibrium~\cite{Evans,Evans2,Wang,Garnier}.

In this work, within the flutuational electrodynamics framework introduced by Rytov we derive the second-order statistical properties of thermal field radiated by a hot body. First, we show that, in the far-field regime, the standard deviation of the radiative heat flux is of 
the same order of magnitude as the mean value, a well-known result from the blackbody theory~\cite{MandelWolf}. On the contrary, in the
near-field regime we find that the standard deviation of the radiative heat flux can be much higher than the mean value, although
the mean value itself is in this regime orders of magnitude larger than the blackbody value. We demonstrate that this significant enhancement of the fluctuation amplitude can be observed  when the medium supports surface polaritons~\cite{JoulainSurfSciRep05}. Finally, we establish that in presence of such waves the correlation time {CT} of PV is of the same order as the relaxation time of atomic vibrations (phonons)  that is much larger than the CT of blackbody radiation. {We further show that for metals the amplitude of fluctuations can also be large, whereas the CT is in the far- and near-field regime of the same order as that of blackbody radiation.}

\begin{figure}
  \epsfig{file = 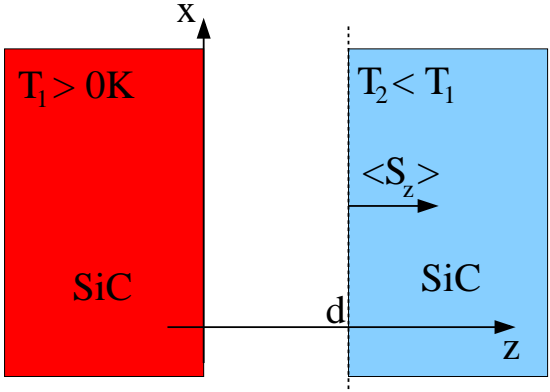, width = 0.4\textwidth}
  \caption{Sketch of the considered configuration: A SiC halfspace at temperature $T_1$ exchanges heat by heat radiation with a second SiC halfspace with $T_2 < T_1$ in a distance $d$.}  \label{Fig:configurations}
\end{figure}

Let us start with the z-component of the mean PV describing the thermal radiation of a semi-infinite medium at temperature $T_1$ into another semi-infinite medium $T_2 = 0\,{\rm K}$ as sketched in Fig.~\ref{Fig:configurations} is given by
\begin{equation}
{
\begin{split}
  \langle S_{1,z} \rangle (\mathbf{r}_d) &= \langle E_{1,x} (\mathbf{r}_d,t) H_{1,y} (\mathbf{r}_d,t)\rangle \\ &\quad- \langle E_{1,y} (\mathbf{r}_d,t) H_{1,x} (\mathbf{r}_d,t)\rangle.
\end{split}}
\end{equation}
Here the index {$1$} symbolizes the fact that the fields are generated by the thermal sources in medium 1 and the brackets denote the ensemble average. {The correlation functions (CFs) are evaluated at the interface of the second medium at $\mathbf{r}_d = (0,0,d)^t$ where the energy transfer to the second body really occurs, and at a given time $t$. Note that we are here considering a non-equilibrium steady-state situation so that the above CFs and the mean PV do not depend on time.} 
In order to determine the second moment we can exploit the Gaussian property of the
thermal fields which allows us to express the higher moments of the fields in terms of 
the second moments~{\cite{Rytov}}. {The main assumption here is that the fluctuational field is composed of a multitude of microfields created by charge and current fluctuations from different volume elements of the medium which give similar and statistically independent contributions. The Gaussian property then follows from the central-limit theorem~\cite{Rytov,SanchezEtAl2004}.} Furthermore, from the rotational symmetry we 
have $\langle E_{1,x}^2 \rangle = \langle E_{1,y}^2 \rangle$ and $\langle H_{1,x}^2 \rangle = \langle H_{1,y}^2 \rangle$;
the components of the electric and magnetic fields are uncorrelated~\cite{MandelWolf} so that
$ \langle H_{1,x} H_{1,y} \rangle = \langle E_{1,x} E_{1,y} \rangle = 0$ and $\langle E_{1,y} H_{1,y} \rangle = \langle E_{1,x} H_{1,x} \rangle = 0$.
Finally, the mixed CFs have the symmetry $ \langle E_{1,x} H_{1,y} \rangle = - \langle E_{1,y} H_{1,x} \rangle$ {\cite{SupplM}}.
With these relations together with the Gaussian property of fields we obtain 
\begin{equation}
  \langle S_{1,z}^2 \rangle = 2 \langle E_{1,x}^2 \rangle \langle H_{1,x}^2 \rangle + \frac{3}{2}\langle S_{1,z} \rangle ^2
\end{equation}
so that the variance {$\langle \bigl( \Delta S_{1,z} \bigr)^2 \rangle \equiv \langle S_{1,z}^2 \rangle - \langle S_{1,z} \rangle^2$} of the normal component  of PV reads
\begin{equation}
    \langle \bigl( \Delta S_{1,z} \bigr)^2 \rangle = 2 \langle E_{1,x}^2  \rangle \langle H_{1,x}^2 \rangle+\frac{1}{2} \langle S_{1,z} \rangle^2.
\label{Eq:Variance}
\end{equation}
Obviously, this quantity is of the same order as the mean heat flux squared and by virtue
of the first term on the right hand side it contains in general contributions from the electric and magnetic fields 
as well. As the normalized standard deviation is concerned it reads accordingly
\begin{equation}
  \sigma_S \equiv \sqrt{\frac{ \langle \bigl( \Delta S_{1,z} \bigr)^2 \rangle}{\langle S_{1,z} \rangle^2}} = \sqrt{\frac{1}{2} + 2 \frac{\langle E_{1,x}^2  \rangle \langle H_{1,x}^2 \rangle}{\langle S_{1,z} \rangle^2}}.
\label{Eq:standardhalfspace}
\end{equation}
Hence, we see that the standard deviation of the thermal emission of a semi-infinite medium is given by the mean value of PV and the electric and magnetic part of the mean energy density.
Expression (\ref{Eq:standardhalfspace}) can of course
also be used to evaluate the standard deviation of the PV for a halfspace emitting into vacuum at $0\:K$ by replacing the permittivity of the right halfspace (i.e. $z > d$)  by that of vacuum. In this case, the mean PV in the expression for the standard deviation will contain the contribution of propagating waves only, whereas
the term $\langle E_x^2  \rangle \langle H_x^2 \rangle$ also contain the contributions of evanescent waves. {A meaningful result for the fluctuations of the PV in the far-field regime can be obtained by evaluating the term $\langle E_x^2  \rangle$ and $ \langle H_x^2 \rangle$ in the limit $d \rightarrow \infty$ or $d \gg \lambda_{\rm th}$.}

In order to evaluate the standard deviation, we need to introduce the CFs of fields at arbitrary separation distances. This can be done in the
 the framework of the theory of fluctuational electrodynamics. To this end, we consider two halfspaces as sketched 
in Fig.~\ref{Fig:configurations} (of permittivity $\epsilon_1 = \epsilon_2 = \epsilon_{\rm SiC}$) separated by a vacuum gap of width $d$ having 
the temperatures $T_1 \neq 0\,{\rm K}$ and $T_2 = 0\,{\rm K}$. In this case the mean value of the PV in z-direction is given by 
 ${ \langle S_{1,z} (\mathbf{r}_d) \rangle = 2 \langle E_{1,x}(\mathbf{r}_d, t) H_{1,y}(\mathbf{r}_d,t) \rangle. }$
From the relation between the fields and the current density and using the fluctuation-dissipation theorem the CFs of electric and magnetic fields read~{\cite{SupplM,Agarwal1975,Eckhardt1978,Eckhardt1979,Doro2011}}
\begin{align}
   \langle E_{1,x}(t)  H_{1,y}(t') \rangle &= \int_0^\infty \!\! \frac{\rd \omega}{2 \pi} \Theta_1 (\omega) \int\!\!\frac{\rd \kappa}{2 \pi} \, \kappa  \frac{\gamma_1' \re^{-2 \gamma_0'' d}}{2 |\gamma_1|^2}  \nonumber\\
                          &\quad\times \biggl( \frac{|t_\rs|^2 |1 + r_\rs|^2}{|D_\rs|^2} \Re\bigl(\gamma_1^* \re^{- \ri \omega \tau} \bigr) \\
                          &\quad          +  \frac{|t_\rp|^2 |1 - r_\rp|^2}{|D_\rp|^2} \frac{|\gamma_1|^2 + \kappa^2}{|k_1|^2 |\epsilon_1|} \Re\bigl(\gamma_1 \epsilon_1^* \re^{- \ri \omega \tau}\bigr) \biggr) \nonumber \\
  \langle E_{1,x}(t)  E_{1,x}(t') \rangle &= \int_0^\infty \!\! \frac{\rd \omega}{2 \pi} \mu_0 \omega \Theta_1 (\omega) \cos(\omega \tau) \int\!\!\frac{\rd \kappa}{2 \pi} \, \kappa\nonumber\\
                          &\quad\times  \frac{\gamma_1' \re^{-2 \gamma_0'' d}}{2 |\gamma_1|^2} 
                           \biggl( \frac{|t_\rs|^2}{|D_\rs|^2} \bigl| 1 + r_s \bigr|^2 \\
                           &\qquad          +  \frac{|t_\rp|^2}{|D_\rp|^2} \frac{|\gamma_1|^2 + \kappa^2}{|k_1|^2} \frac{|\gamma_1|^2}{|k_1|^2} \bigl| 1 - r_p \bigr|^2 \biggr) \nonumber, \\
  \langle H_{1,x}(t)  H_{1,x}(t') \rangle &= \int_0^\infty \!\! \frac{\rd \omega}{2 \pi} \epsilon_0 \omega \Theta_1 (\omega) \cos(\omega \tau) \int\!\!\frac{\rd \kappa}{2 \pi} \, \kappa \nonumber \\
                        &\quad\times \frac{\gamma_1' \re^{-2 \gamma_0'' d}}{2 |\gamma_1|^2} \biggl( \frac{|t_\rs|^2}{|D_\rs|^2} \frac{|\gamma_1|^2}{k_0^2} \bigl| 1 + r_s  \bigr|^2 \\
                           &\qquad          +  \frac{|t_\rp|^2}{|D_\rp|^2} \frac{|\gamma_1|^2 + \kappa^2}{k_0^2}  \bigl| 1 - r_p  \bigr|^2 \biggr) \nonumber,
\end{align}
with $\tau := t - t'$ and the mean energy of a harmonic oscillator given by  $ \Theta_1 (\omega) = \hbar \omega/ \bigl( \exp(\hbar \omega / \kb T_1)- 1\bigr)$. {Here} $D_{\rs/\rp} = |1 - r_{\rs/\rp}^2 \re^{2 \ri \gamma_0 d}|^2$, $\gamma_1^2 = k_0^2 \epsilon_1 - \kappa^2$, $\gamma_0^2 = k_0^2 - \kappa^2$, $k_1 = \sqrt{\epsilon_1} k_0$ and $k_0 = \omega/c$; $t_{\rs/\rp}$ and $r_{\rs/\rp}$ are the Fresnel transmission and reflection coefficients of the single interface;{$\epsilon_0$ and $\mu_0$ are  the permittivity and permeability of vacuum.}

With these relations we can determine the fluctuations of  PV between any couple of
isotropic and homogeneous halfspaces considering only the thermal radiation from a single halfspace with $T_1 \neq 0\,{\rm K}$. 
In particular it is possible to derive from these expressions the moments of heat flux radiated by a blackbody of temperature $T_1$   in vacuum. Indeed, in this case by setting the permittivity of the materials $\epsilon_1$ to that of vacuum, i.e.\ $\epsilon_1 \equiv 1$ so that $t_\rs=t_\rp=1$ and $\gamma_1=\gamma_0$  
then it is easy to see that~\cite{SupplM}
\begin{equation}
  \langle S_z \rangle \equiv \langle S_{\rm BB,z} \rangle  = \sigma_{\rm BB} T_1^4,  \label{Eq:SzBB} 
\end{equation}
{and}
\begin{align}
\langle  E^2_x (t) \rangle &\equiv \langle E_{\rm BB,x}^2\rangle   = c \mu_0 \frac{2}{3} \sigma_{BB} T_1^4 , \label{Eq:ExBB} \\
\langle  H^2_x (t) \rangle &\equiv \langle H_{\rm BB,x}^2\rangle   = c \epsilon_0 \frac{2}{3} \sigma_{BB} T_1^4  \label{Eq:HxBB}
\end{align}
introducing the Stefan-Boltzmann constant $\sigma_{\rm BB}$.
It follows that the normalized standard deviation for the blackbody radiation reads
\begin{equation}
  \sigma_{S,\rm BB} = \sqrt{\frac{25}{18}} \label{Eq:sigmaBB}
\end{equation}
showing that the standard deviation of PV is of the same order as its  mean value. 
This result is obviously consistent with the well-known deviation $\sigma = \langle I \rangle /\sqrt{2}$  of unpolarized thermal radiation, $\langle I \rangle$ being the mean value of the intensity~\cite{MandelWolf}.

\begin{figure}
  {\epsfig{file = 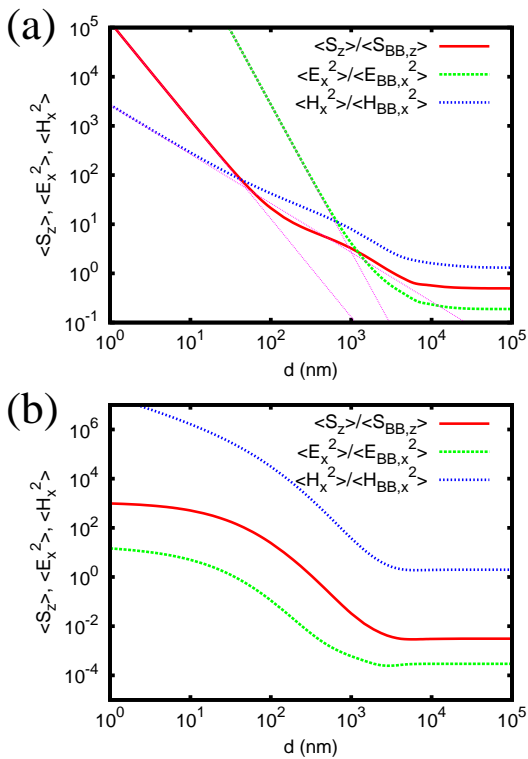, width = 0.4\textwidth}}
  \caption{$\langle S_z \rangle$, $\langle E_x^2 \rangle$ and $\langle H_x^2 \rangle$ as function of gap size $d$ for {two (a) SiC and (b) Au} halfspaces with $T_1 = 300\,{\rm K}$ and $T_2 = 0\,{\rm K}$; all quantities are normalized to the blackbody values given in Eq.~(\ref{Eq:SzBB}), (\ref{Eq:ExBB}) and (\ref{Eq:HxBB}) for $T_1 =  300\,{\rm K}$. The thin lines {in (a)} are the quasi-static results showing that $\langle S_z\rangle \propto 1/d^2$, $\langle E_x^2 \rangle \propto 1/d^3$ and $\langle H_x^2 \rangle \propto 1/d$ in the strong near-field regime. \label{Fig:SEEHH}}
\end{figure}

Now let us pay attention to heat exchanges between two bulk samples made of silicon carbide (SiC) a polar material whose permittivity at frequency $\omega$ can be described by the Drude-Lorentz model {and two samples made of gold (Au) described by the Drude model~\cite{Palik} (see also~\cite{SupplM}).}
We first show in Fig.~\ref{Fig:SEEHH} plots of CFs as derived above and normalized by
the CFs for a blackbody. {For SiC it} can be seen that in the quasi-static limit $\langle S_z \rangle \propto 1/d^2$, $\langle E_x^2 \rangle \propto 1/d^3$,  and $\langle H_x^2 \rangle \propto 1/d$ due to the near-field contribution. {These distance dependences are universal features in the quasi-static limit. For Au all the curves would have the corresponding distance dependences for $d \rightarrow 0$ (see~\cite{SupplM}), but for the shown values of $d$ the quasi-static regime is not yet fully reached.}

\begin{figure}
  {\epsfig{file = 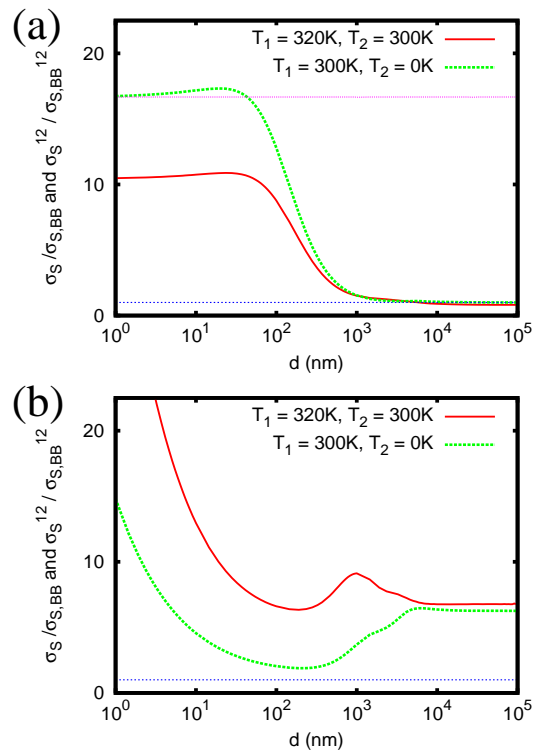, width = 0.4\textwidth}}
  \caption{Normalized standard deviation $\sigma_S$ from Eq.~(4) and Eq.~(14) as function of the gap size $d$ for two {(a) SiC and (b) Au} halfspaces with $T_1 = 300\,{\rm K}$ and $T_2 = 0\,{\rm K}$ and $T_1 = 320\,{\rm K}$ and $T_2 = 300\,{\rm K}$.  The standard deviation is normalized to the blackbody results $\sigma_{S,{\rm BB}} \approx 1.18$ from Eq.~(\ref{Eq:sigmaBB}) and $\sigma_{S,{\rm BB}}^{12} \approx 6.25$ from Eq.~(\ref{Eq:BBzwei}), resp. {The vertical lines are the quasi-static limits~\cite{SupplM} and the blackbody value.} 
  }
 \label{Fig:sigmaSSiC}
\end{figure}

{The standard deviation $\sigma_S$ shown in Fig.~\ref{Fig:sigmaSSiC}(a) for SiC and in Fig.~\ref{Fig:sigmaSSiC}(b) for Au is $d$ independent in the far-field regime (i.e. $d \gg \lambda_{\rm th}$) as can be expected from the fact that the CFs are $d$ independent in this case. Since SiC is a very good absorber in the infrared it is not surprising that $\sigma_S$ is very close to the $\sigma_{\rm S,BB}$. In the near field regime $\sigma_S$ increases and converges to a constant value in the quasi-static limit~\cite{SupplM}. On the other hand, for Au  $\sigma_S$ is relatively large in the far-field regime and first decreases when making $d$ smaller and then increases for very small distances. The value of $\sigma_S$ would converge to its quasi-static limit~\cite{SupplM} for $d \rightarrow 0$. Note that this convergence to a distance independent value for $d \rightarrow 0$ is a universal feature, whereas the value to which $\sigma_S$ converges depends on the material properties and in particular on the losses~\cite{SupplM}. For SiC we find the quasistatic limit $\sigma_S \approx 16.7\times\sigma_{S,{\rm BB}} \approx 20$ and for Au we find $\sigma_S \approx 2274\times\sigma_{S,{\rm BB}} \approx 2683$. The fluctuational amplitude is therefore for metals potentially higher. However, at $d=10\,{\rm nm}$ for SiC the standard deviation is about $20 \times \langle S_z \rangle$, whereas for Au it is about $5 \times \langle S_z \rangle$. The fluctuations do therefore rapidely increase due to the near-field enhanced heat flux and local density of states, which is a result of the contribution of the surface phonon polaritons in SiC and eddy currents in Au~\cite{JoulainSurfSciRep05}.}

\begin{figure}
  \epsfig{file = 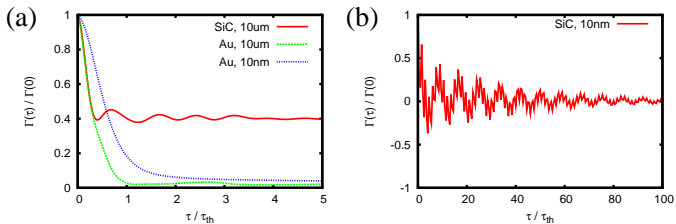, width = 0.5\textwidth}
  \caption{{Normalized temporal CFs $\Gamma(\tau) = \langle S_z(t) S_z(t + \tau)\rangle$ for two SiC and Au halfspaces as function of $\tau$ normalized to $\tau_{\rm th} = \hbar/ \kb T_1 \approx 2.5\times10^{-14}\,{\rm s}$ at $T_1 = 300\,{\rm K}$ and $T_2 = 0\,{\rm K}$ in far-field ($10\,\mu{\rm m}$) and near-field ($d = 10\,{\rm nm}$) regime.}    \label{Fig:correlationtime}}
\end{figure}

Finally, in the general situation where $T_2 \neq 0\,{\rm K}$, which means that also the thermal sources in the second halfspace need to be taken
into account, one can again in a similar way derive the variance {$ \langle \bigl( \Delta S_z^{12} \bigr)^2 \rangle \equiv \langle (S_{1,z} - S_{2,z})^2 \rangle - \langle (S_{1,z} - S_{2,z}) \rangle^2 $} of the heat
flux. Furthermore, assuming that the fluctuational sources in the two bodies and also the generated fluctuating fields are 
uncorrelated, we obtain the general expression
\begin{equation}
\begin{split}
  \langle \bigl( \Delta S_z^{12} \bigr)^2 \rangle                       &= \frac{1}{2} \langle S_{1,z} \rangle^2 + 2 \langle E_{1,x}^2  \rangle \langle H_{1,x}^2 \rangle \\
                                             &\quad + \frac{1}{2} \langle S_{2,z} \rangle^2 +  2 \langle E_{2,x}^2  \rangle \langle H_{2,x}^2 \rangle,
\end{split}
\label{Eq:Variance2}
\end{equation}
where $\langle S_{2,z} \rangle$, $\langle E_{2,x}^2  \rangle$ and $ \langle H_{2,x}^2 \rangle$ take a similar form as 
$\langle S_{1,z} \rangle$, $\langle E_{1,x}^2  \rangle$ and $ \langle H_{1,x}^2 \rangle$ but with $T_2$ instead of $T_1$. Since we assume the absence of correlation between the sources of two different media, we find that the fluctuations are additive. {The} relative standard deviation {is}
\begin{equation}
  \sigma_S^{12} =  \frac{\sqrt{\langle \bigl( \Delta S_z^{12} \bigr )^2 \rangle}}{\langle S_{1,z}\rangle - \langle S_{2,z}\rangle}.
\end{equation}
From this expression it becomes clear that this deviation is larger than in the case where $T_2 = 0\,{\rm K}$. As {before} we can derive the result for two blackbodies in interaction 
 \begin{equation}
  \sigma_{S, BB}^{12} =  \frac{\sqrt{\frac{T_1^8}{T_2^8} + 1}}{\frac{T_1^4}{T_2^4} - 1} \sqrt{\frac{25}{18}}.
\label{Eq:BBzwei}
\end{equation}
Furthermore, it should be noted that in the limit $\Delta T = T_1 - T_2 \rightarrow 0$ the variance in (\ref{Eq:Variance2}) converges to 
a constant which is, due to the additivity,  just twice the value given by eq.~(\ref{Eq:Variance}) corresponding to the deviation for a single semi-infinite medium. That means, although the mean heat flux becomes zero in this limit, the fluctuations of  heat flux persist. Therefore, the relative standard deviation  $\sigma^{12}_S$ can be very large for small temperature differences and even diverges when $\Delta T \rightarrow 0$ as can be nicely seen from expression for the blackbody case where $\sigma^{12}_{S,BB}=\frac{5}{12} \frac{T_1}{\Delta T}$ for small $\Delta T$. In Fig.~\ref{Fig:sigmaSSiC} we find {at d=10~nm} for the heat flux between two SiC {(Au)} halfspaces a relative standard deviation of $\sigma_{S}^{12} \approx 65$ {(81)} times the measured heat flux value, which is large and should be measurable in existing near-field heat flux experiments. 

We have seen that the heat flux fluctuations are large. But, in order to assess on what extent these fluctuation are measurable it is important to evaluate on which time scale these fluctuations happen. From the blackbody theory
it is well known that the CT of thermal field  is on the order of $\tau_{\rm th} = \hbar / \kB T $ that is about $  2.5\times10^{-14}\,{\rm s}$ at $T = 300\,{\rm K}$. This timescale is very similar to the CT we observe in Fig.~\ref{Fig:correlationtime}(a) by plotting the temporal CF {$\Gamma (\tau) = \langle S_z(t) S_z(t + \tau) \rangle$ given by}~{\cite{SupplM}}
\begin{equation}
\begin{split}
         \Gamma (\tau) &= 2 \langle E_x(t)E_x(t + \tau)\rangle  \langle H_x(t)H_x(t + \tau)\rangle \\
                       &\qquad+ 2 \langle E_x(t) H_y(t + \tau) \rangle^2 + \langle S_z (t) \rangle^2
\end{split}
\end{equation}
 of the heat flux between two SiC {and Au} halfspaces as function of $\tau = t' - t$ in the far-field at a distance of $d = 10\,\mu {\rm m}$. 
Although the time scale of  $\tau_{\rm th}$ is extremely small, this temporal correlation has been measured in the context of photon 
bunching~\cite{Morgan, TanEtAl2014}. In contrast,
if we plot {$ \Gamma (\tau)$} for a near-field distance of $d = 10\, {\rm n m}$ in Fig.~\ref{Fig:correlationtime} (b)
we can observe that the timescale on which the heat flux is temporarily correlated is about 50$\times \tau_{\rm th} = 1.25\times10^{-12}\,{\rm s}$ {due to the quasi-monochromatic contribution of surface-phonon polariton~\cite{ShchegrovEtAl2000}. On the other hand, for Au the CT does not change much in the near-fied regime as can be seen in Fig.~\ref{Fig:correlationtime} (a).}  Hence the timescale of fluctuations of the radiative heat flux 
in the near field {can be} on the same order of magnitude as that of {relaxations} of the phonons in a medium~{\cite{LindeEtAl1980}}. 

In conclusion, we have introduced a general theory to describe fluctuations of radiative heat flux exchanged between two  bodies. We have shown that at subwavelength distances large fluctuations of heat flux can be observed when heat exchanges results from surface polariton coupling. This is in huge contrast to the findings of the zero-point fluctuations of Casimir force~\cite{Barton1991,MessinaPassante2007}. We think that this theory should allow for testing the Crook fluctuation theorem~\cite{Evans,Evans2,Reid,Evans3,Jarzynski}. Hence, by measuring the time evolution of heat flux exchanged between two nanostructures it is in principle possible to calculate the probability to observe an instantaneous negative flux transferred from a cold body to a hot one and to compare this value with the probability of a transfer in the opposite direction.  Beyond this fundamental test, this theory can be used to investigate the irreversibility mechanisms associated to thermal photon exchanges~\cite{Reid,Evans3,Cohen} or to explore the performances of nanomachines such as Brownian motors.

%
%

S.-A. Biehs acknowledges discussions with Andreas Engel, Achim Kittel (Oldenburg University) and Riccardo Messina (CNRS). { P. B.-A.  acknowledges discussions with Miguel Rubi (Barcelona University).}

%
%

\end{document}